# Sagnac interferometer-based noise-free superresolution using phase-controlled quantum erasers

Byoung S. Ham[1,2]
[1]Department of Electrical Engineering and Computer Science, Gwangju Institute of Science and Technology, 123 Chumdangwagi-ro, Buk-gu, Gwangju 61005, South Korea
[2]Qu-Lidar, 123 Chumdangwagi-ro, Buk-gu, Gwangju 61005, South Korea
(March 15, 2025; bham@gist.ac.kr)

**Abstract**
Interferometer-based precision measurements have been intensively studied for sensing and metrology over the past half century. In classical optics, the resolution and phase sensitivity of an optical signal are confined by diffraction limit and shot-noise limit (SNL), respectively. Highly entangled photon pairs, i.e., N00N states have been adapted to overcome SNL in quantum sensing over the last two decades. Recently, coherent light-excited quantum sensing has also been proposed and demonstrated for macroscopic quantum sensing to overcome the limited N scalability in N00N-based quantum sensing. Here, a Sagnac interferometer-based superresolution is proposed to solve environmental noises inevitable in an interferometer. Furthermore, a spatial light modulator takes over the role of phase-controlled quantum erasers to solve the linear optics-based complexity issue in the coherently-excited superresolution. Thus, the proposed Sagnac superresolution can beat the state-of-the-art ring laser gyroscope applied for inertial navigation and geodesy.

**Introduction**
Precision measurements have been a long lasting goal in sensing and metrology for both science and engineering ranging from imaging [1-5] to gravitational wave detection [6,7]. The main stream of precision measurements is interferometer-based optical sensing [8-11], where the ultimate goal is to overcome the diffraction limit in a single slit (or an aperture) and the shot-noise limit (SNL) in the Fisher information. A multi-slit based optical sensing has become a general method to overcome the diffraction limit toward high resolution via many-wave interference [12,13]. This slit number-proportional resolution enhancement also contributes to enhance phase sensitivity [13-15]. On the other hand, the phase sensitivity in an interferometer can be further enhanced via a multiple readout-based intensity product expressed by Fisher information [16,17]. With a normal distribution of measurement events, the minimum error or phase sensitivity is known as SNL in classical physics [14]. An optical cavity is a common tool for the resolution enhancement via multi-wave interference, where the enhancement is proportional to the slit number or Q factor of an optical cavity [12,18]. So far, the enhanced resolution in a multi-slit or an optical cavity system has been well applied to off-the-shelf wavelength meter and ring laser gyroscope (RLG), whose resolution enhancement factor is beyond $10^4$ over the diffraction limit [18].

Quantum sensing and metrology have been intensively studied using N00N states in a Mark-Zehnder interferometer (MZI) over the last two decades to overcome the diffraction limit and SNL [14,15]. Like SNL demonstrated by projection measurements of the MZI output signal for higher-order intensity product [17], the projection measurement is a major tool for the N00N-based quantum sensing [19-21]. As a result, superresolution [20-23] and supresensitivity [23] have been demonstrated with N00N states in MZI for quantum sensing. The superresolution is rooted in photonic de Broglie wavelength (PBW) $\lambda_B$, whose resolution enhancement is linearly proportional to the intensity-product order N of the N00N state, resulting in $\lambda_B = \lambda_0/N$ [22]. Thus, the superresolution is different from the multi-slit or cavity-enhanced resolution in classical physics due to no $\lambda_B$ [12,16-18]. So far, the major hurdle of the N00N-based quantum sensing is limited N with less than 20 [24]. By the Nth-order intensity correlation of the N00N-based superresolution, the supersensitivity with $\sqrt{N}$ quantum gain is self-contained if a near perfect visibility of PBW is satisfied [23]. To beat the diffraction limit with more than $10^4$-times enhanced resolution, the minimum N of a N00N state in quantum sensing should be $N > 10,000$, which has never been reported yet.

Recently, coherently-excited macroscopic superresolution has been proposed [25] and experimentally demonstrated in MZI [26]. A phase-controlled intensity product has already been demonstrated for the coherent superresolution in a single photon regime using a beam splitter (BS) [27,28] and MZI [25,26]. Unlike the BS cases [27,28], however, the superresolution of MZI case [25,26] is compatible with the N00N-based quantum sensing [14,19-23] in terms of measurement-based PBW. The same idea has also extended to a Michelson interferometer [29], targeting for radar applications. For the N00N-based quantum sensing, a projection measurement method has been applied to an MZI output port to block unwanted intensity-product order [19-21]. In the macroscopic superresolution, input laser light is polarization controlled [30,31], as in the N00N-based quantum sensing [21], for the macroscopic quantum eraser [25,26]. To solve the environmental noise constraint resulting from mechanical vibrations, temperature fluctuations, and air turbulences, here, the MZI is replaced by a Sagnac interferometer for noise-free Sagnac superresolution. Due to the macroscopic feature [25,26,29], thus, the proposed Sagnac superresolution gives a solution to the N limited quantum sensing to beat the classical counterpart.

**Results**

*A. Background: polarization-basis projection measurements*

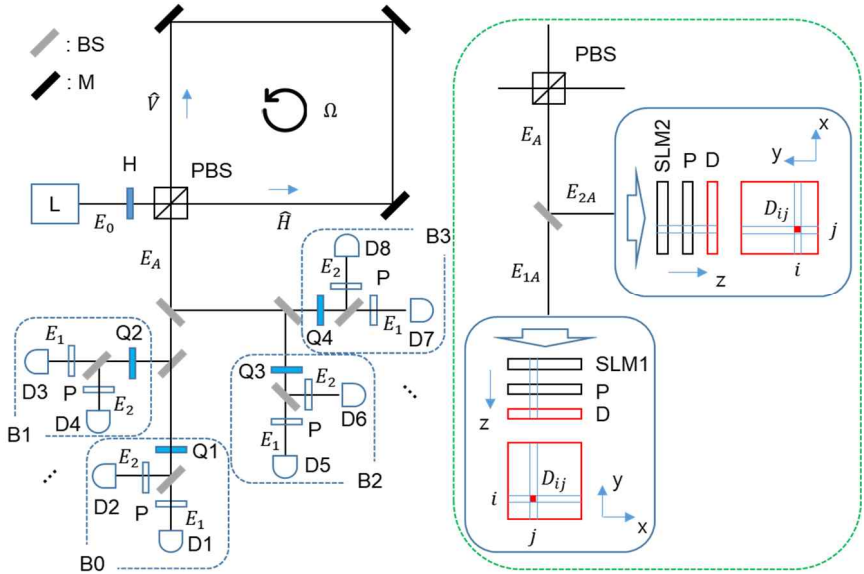

**Figure 1.** Schematic of noise-free superresolution in a Sagnac interferometer. (Inset) SLM-based 2D polarization-basis projection measurements. L: laser, H: 22.5°-rotated half-wave plate, PBS: polarizing beam splitter, Q: quarter-wave plate, P: polarizer, D: photodetector, SLM: spatial light modulator. Ω: angular velocity of the system.

Figure 1 shows a schematic of the proposed noise-free Sagnac superresolution using phase-controlled projection measurements. By PBS, the output field $E_A$ of the Sagnac interferometer is set for a near-perfect throughput into one output port with no interference fringes. For the proposed Sagnac superresolution, the output light $E_A$ is equally divided into K optical blocks (see the dotted squares) and phase controlled for the projection measurements by QWPs [25,26,29]. For this projection measurement, orthogonal polarization bases $(\hat{H}; \hat{V})$ of an input light are provided by the 22.5-degree rotated half-wave plate (HWP), H. Then, the measured optical signals on detectors Ds show interference fringes, satisfying the coherently excited quantum eraser [30,31]. This quantum eraser has no discrepancy between single photons and continuous-wave (CW) light, satisfying the macroscopic quantum feature. The maximum number K of the optical blocks denoted by dotted squares, corresponding to K sets of quantum erasers, is limited by the photon number of the output field $E_A$.

Thus, the intensity-product order of the proposed Sagnac superresolution is practically unlimited due to the huge number of photons a laser. Thus, each block behaves as a pair of quantum erasers having opposite fringes [25,26,29], where the fringes of all blocks are equally shifted in a phase space, as shown in Fig. 2. Here, QWP is a birefringent material to induce a phase sift $\xi$ between ordinary ($\hat{H}$) and extraordinary ($\hat{V}$) axes [12,29]. This phase shift $\xi$ is individually controllable from block to block in Fig. 1 for the noise-free Sagnac superresolution (see *Section B*

The rotation angles of all QWPs are set to be discretely assigned to satisfy equally spaced fringe shifts between all quantum erasers in Fig. 1 (see Fig. 2). These fringe-shifted quantum erasers are for the first-order intensity correlation of the divided output fields and plays the key role in the proposed Sagnac superresolution, as demonstrated in a single-photon regime of MZI [26,27]. Due to the perfect spatial overlap between orthogonally polarized counterpropagating lights in the Sagnac interferometer, any environment-caused phase fluctuations resulting from mechanical vibrations, temperature fluctuations, and air turbulences are automatically and completely cancelled out in $E_A$ due to sharing the same path. Thus, the only source of the phase-difference control is $\zeta$ caused by the rotation velocity $\Omega$, which is known as the Sagnac effect [32,33]. As a result, the proposed noise-free Sagnac superresolution is rooted in the phase-controlled polarization-basis projection measurements via Nth-order intensity product between K optical blocks: N=2K. For the scalability of the intensity-product order N in the Sagnac superresolution, a pair of SLMs replaces the linear optics blocks composed of QWPs and BSs to solve the complexity issue (see Inset).

B. *Analysis: phase-controlled polarization-basis projection*

The polarization-basis projection in Fig. 1 is coherently analyzed for the general solution of the proposed noise-free Sagnac superresolution. The amplitude $E_A$ of the output field from the Sagnac interferometer is represented by:

$$\boldsymbol{E}_A = \frac{E_0}{\sqrt{2}}(\hat{H} - \hat{V}e^{\zeta}), \tag{1}$$

where $E_0$ is the amplitude of the input light just before entering the Sagnac interferometer. For simplicity, loss-free linear optics is assumed as usual. The Sagnac effect-caused phase shift $\zeta$ originates in the time delay between counterpropagating lights due to the angular velocity $\Omega$ [32,33]. Each pair of quantum erasers in each block as shown in the dotted box is deterministically phase controlled by QWPs, resulting in the corresponding phase shift $\xi_k$ in the $\hat{V}$ component [25,29]. The phase control $\xi_k$ in the kth block is set for $2\pi k/N$, where N=2K is the total number of divided fields [29]. The role of a 45°-rotated polarizer P is for the polarization-basis projection measurements, enabling the quantum eraser [30]: Without Ps, no interference fringe occurs due to the distinguishable photon characteristics of $\hat{H}$ and $\hat{V}$ in the Sagnac interferometer. The rotation angle of QWP is set by the fast axis aligned to the vertical direction, resulting in a $\pi/2$ phase shift with respect to $\hat{H}$ component [12]. Without QWP, no phase shift to the $\hat{V}$ component is induced. If the rotation angle is 90°, i.e. along the fast-axis horizontal, then a π-phase shift is induced to the $\hat{V}$ component [12].

From Eq. (1), thus, the following relations are obtained for the kth optical block in Fig. 1:

$$\boldsymbol{E}_{k1} = \gamma_{k1}E_0(\sqrt{2})^{-N/2}\big(H\cos\theta - (-1)^k V\sin\theta\, e^{i(\zeta-\xi_k)}\big)\hat{p}, \tag{2}$$

$$\boldsymbol{E}_{k2} = \gamma_{k2}E_0(\sqrt{2})^{-N/2}\big(H\cos\theta + (-1)^k V\sin\theta\, e^{i(\zeta-\xi_k)}\big)\hat{p}, \tag{3}$$

where $\hat{p}$ is the optical axis of the polarizer, and $\gamma_k$ is a *k*-dependent global phase. The subscript '*j*' in $\boldsymbol{E}_{kj}$ indicates the field number in $E_j$ for measurements. The controlled phase in each block is denoted by $\xi_k$, where $k = 0, 1, \ldots, \frac{N}{2} - 1$. The notation of $H$ and $V$ in Eqs. (2) and (3) are dummy just to indicate their origins. In Eqs. (2) and (3), the polarization-basis projection onto the polarizer's axis $\hat{p}$ results in indistinguishable photon characteristics via $\hat{H} \to \cos\theta \hat{p}$ and $\hat{V} \to \sin\theta \hat{p}$. Thus, interference fringes are resulted in all detectors, demonstrating the quantum eraser (see below) [30,34]. By BS in each block, the reflected light experiences a $\pi/2$ phase shift with respected to the transmitted [35]. At the same time, the BS reverses the polarization direction of $\hat{H}$, as in the mirror images [12,25]. Based on the same birefringence material of SLM as QWP, the same role of phase control can be conducted, where each QWP-based optical block in Fig. 1 corresponds to each SLM pixel pair in the Inset. The pixel-to-pixel phase control of light using SLM is well matured [36].

Generalized solutions of corresponding mean intensities for Eqs. (2) and (3) are derived as follows for $\theta = 45°$ of all Ps:

$$\langle I_{k1}\rangle = I_0 2^{-K}\langle 1 - (-1)^k \cos(\zeta - \xi_k)\rangle, \tag{4}$$

$$\langle I_{k2}\rangle = I_0 2^{-K}\langle 1 + (-1)^k \cos(\zeta - \xi_k)\rangle, \tag{5}$$

where $I_j = E_j E_j^*$ and $\xi_k = \frac{\pi k}{N}$. The K is the light-split level for N divided lights. The first (second) subscript in $I_{ab}$ indicates the block (optical field) number. Equations (4) and (5) represent a pair of quantum erasers in the kth block, resulting in interference fringes [29,30]. In each block, the intensity product between Eqs. (4) and (5) show twice increased fringe numbers, as shown in Fig. 2(c). To satisfy PBWs for all higher-order (N>2) intensity products, the key requirement is to properly set $\xi_k$ by kth QWP for an equal fringe shift between $I_j$s [26,29]. This condition is $\xi_k = 2\pi k/N$, in which each pair of quantum erasers in each block is preset by BS for the out-of-phase relation [29]. Thus, a $\pi$-modulus $\xi_k$ ($\xi_{N/2} - \xi_1 = \pi$) becomes the most important condition of the phase control in the proposed Sagnac superresolution. Then, the Nth-order intensity product between K blocks in Fig. 1 satisfies N-times increased fringe numbers, corresponding to the PBW-like superresolution in the macroscopic regime (see Fig. 2). Such superresolution has been demonstrated in MZI theoretically [25] and experimentally [26]. Most importantly, a nearly perfect fringe visibility of the quantum erasers is satisfied [26,30], satisfying both superresolution and superssensitivity in quantum sensing [23,38].

Finally, the second-order intensity product $R_{12}^{(2)}$ for an arbitrary kth block is derived from Eqs. (4) and (5):

$$\langle R_{12}^{(2)}(k)\rangle = I_0^2 2^{-N}\langle \sin^2(\zeta - \xi_k)\rangle. \tag{6}$$

Thus, the general solution of the Nth-order intensity product between all blocks in Fi.g 1 is as follows:

$$\langle R_{12}^{(N)}(\zeta,\xi_k)\rangle = I_0^N 2^{-KN} \prod_{k=0}^{\frac{N}{2}-1}\langle \sin^2(\zeta - \xi_k)\rangle. \tag{7}$$

Using 1 mW power of He/Ne laser, each channel of the N-divided light ports in Eq. (7) contains $10^{15}/N$ photons before P, satisfying a macroscopic regime in intensity correlations even for a millionth order of PBWs with $N = 10^6$, which is easily provided by an off-the-shelf SLM.

*C. Numerical calculations*

Figure 2 shows numerical calculations of the Nth-order (N=512) intensity products in Fig. 1 using Eqs. (4) and (5) as a function of the Sagnac effect $\zeta$ caused by $\Omega$ and QWP-assigned phase control $\xi_k$. Figures 2(a) and (b) represent individual intensities of all 256 blocks, where the block number-dependent control phase $\xi_k$ is applied to the $\hat{V}$ component only by QWPs [25,29,37]. For each $\xi_k$, $I_1$ and $I_2$ are confirmed to satisfy the out-of-phase relation in the fringe patterns of individual detectors of each block for the quantum eraser set. Thus, the intensity product between paired detectors in each block should satisfy doubled fringes by the second order, as shown in Fig. 2(c). Figure 2(d) represents $\xi_k$-dependent second-order intensity products for four blocks (N=8) out of 256 blocks in total in Fig. 2(c), as marked by dashed lines. For the fourth-order (N=4) intensity product satisfying doubled fringes compared to the second-order intensity product in Fig. 2(d), a proper choice of $\xi_k$ $(=\frac{2\pi k}{N})$ for a certain block in Fig. 2(c) is essential.

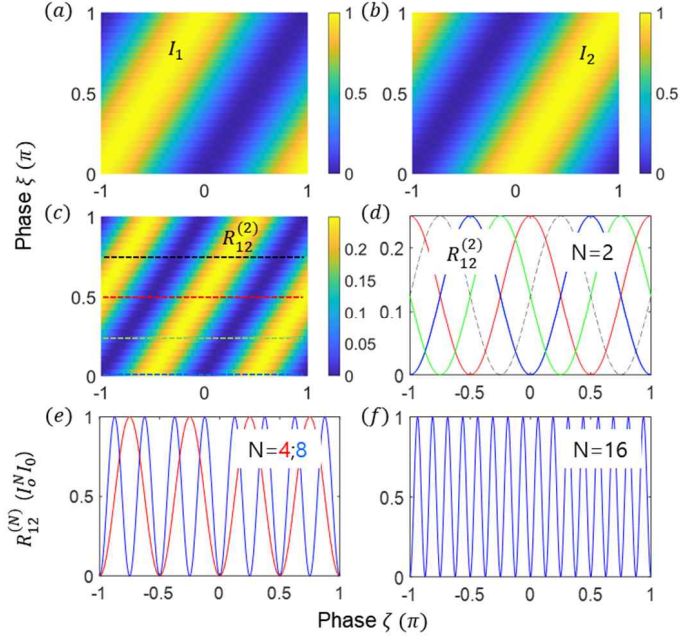

**Figure 2.** Numerical calculations of the intensity product in Fig. 1. (a) and (b) Individual first-order intensity correlation $I_j$. N=512. (c) second-order intensity correlation $R_{12}^{(2)}$. (d) $\xi$-dependent fringe shift. $\xi=0$(B); $\frac{\pi}{4}$(G); $\frac{\pi}{2}$(R); $\frac{3\pi}{4}$ (dotted). (e) N=4(R); 8(B). (f) N=16. $R_{12}^{(2)} = I_1 I_2$, $R_{12}^{(N)} = \prod_{\xi_k} R_{12}^{(2)}(\xi_k)$. $\xi_k = k\pi/N$. Intensity is normalized.

Given configuration with total N, thus, a reduced intensity product can be obtained by choosing equally spaced $\xi_k$s. If it is for N=4, i.e., K=2, $\xi_k = \pi/2$ (red dashed line) is taken in addition to $\xi_k = 0$ (blue dashed line as a reference) in Fig. 2(c). Their intensity products for lower orders (N=2) satisfy the out-of-phase relation between them, as shown in the blue and red curves in Fig. 2(d). Likewise, the green and dashed curves in Fig. 2(d) also satisfies the out-of-phase relation with the set of blue and red curves. In addition, the intensity product fringes between the $\xi_k$-chosen blocks, i.e., the green and dashed curves are equally shifted from those of blue and red curves, respectively (not shown). Thus, the 4th-order intensity product between two paired blocks, corresponding to each pair of curves in Fig. 2(d), results in again doubled fringes, as shown by the red curve in Fig. 2(e). In the same analogy, the blue curve in Fig. 2(e) is for the 8th-order intensity product between all blocks in Fig. 2(d), resulting in eight-times enhanced fringe resolution, satisfying PBW at $\lambda_B = \lambda_0/8$. With the linear scalability of the phase-controlled blocks in Fig. 1, the superresolution can also be extended to N=16 by choosing eight equally spaced $\xi_k$s at a step of $\pi/8$ in Fig. 2(c): Figure 2(f) confirms 16-times enhanced resolution for the proposed Sagnac superresolution in Fig. 1. As numerically demonstrated in Fig. 2, thus, the general solution of Eq. (7) for the PBW-like macroscopic Sagnac superresolution is now numerically confirmed, where the fringe visibility is near perfect owing to the noise-free Sagnac interferometer. This near-perfect fringe visibility also self-contained super-sensitivity [23], resulting in the macroscopic Heisenberg limit overcoming SNL.

To solve the complexity issue of the linear optics-based projection measurements in Fig. 1 [25,26,29], the linear optics blocks are replaced by a pair of SLMs, as shown in the Inset [37]. The SLM pair takes over the role of QWPs for the same phase control of $\xi_k$ in each pair of pixels. The QWP-based phase control $\xi_k$ is now conducted by a synchronized voltage applied to the pair of SLM pixels. For this, Eq. (1) can be rewritten for the kth pixel pair of SLMs as follows:

$$\boldsymbol{E}_k = \frac{E_0}{2}\left(\widehat{H} - \widehat{V}e^{i(\zeta-\xi_k)}\right). \tag{8}$$

The pair of detectors in all blocks through QWPs in Fig. 1 is now replaced by a pair of SLM pixels, where the SLM control voltages to the pixel pair are synchronized for the same phase control $\xi_{k2} = \xi_{k1} = \frac{\pi k}{N}$. Thus, Eq. (8) is rewritten even for the pair of SLM blocks $D_{ij}$ in the Inset of Fig. 1 as $\boldsymbol{E}_{k1} = \frac{E_0}{N}\left(1 - e^{i(\zeta-\xi_k)}\right)\hat{p}$ and $\boldsymbol{E}_{k2} = -i\frac{E_0}{N}\left(1 + e^{i(\zeta-\xi_k)}\right)\hat{p}$, where $\boldsymbol{E}_{1A} = \frac{E_0}{2}\left(\widehat{H} - \widehat{V}e^{i(\zeta-\xi_k)}\right)$ and $\boldsymbol{E}_{2A} = -\frac{iE_0}{2}\left(\widehat{H} + \widehat{V}e^{i(\zeta-\xi_k)}\right)$. For K pixels of SLM, $\boldsymbol{E}_{1A}$ and $\boldsymbol{E}_{2A}$ are divided into K under the ideal beam expansion. Then, the intensities of the kth pair of pixels in SLMs satisfy the same out-of-phase relation, as in Figs. 2(a) and (b), resulting in Eq. (6) for the corresponding second-order intensity product. Unlike the N00N-based superresolution in quantum sensing [14,15,19-23], the N scalability issue is overcome by the total pixels of SLM. The reduction factor $\left(\frac{I_0}{2}\right)^N$ has no critical problem, unless N is less than the photon number of the input laser. With 1 mW, HeNe laser, each SLM pixel contains far more than million photons for $N = 10^6$, additionally satisfying a high signal-to-noise ratio in the fringe visibility as a benefit of the macroscopic superresolution compared to the single photon case.

**Discussion**

From the universal scheme of the proposed noise-free Sagnac superresolution based on a Sagnac interferometer via polarization-basis projection measurements in Fig. 1, the numerical calculations in Fig. 2 for the generalized solution of Eq. (7) can intuitively lead to the form of PBWs:

$$\langle R^{(K)}_{B_0 B_1 \ldots B_k \ldots B_{N/2}} \rangle = \left(\frac{I_0}{2^4}\right)^N \langle \sin^2(N\varphi) \rangle, \tag{9}$$

where $B_k$ indicates the kth phase-controlled quantum eraser block (or kth pixel pair of SLMs). Here, the effective phase $N\varphi$ in Eq. (9) represents PBWs in supreresolution quantum sensing with N00N states [14,15].

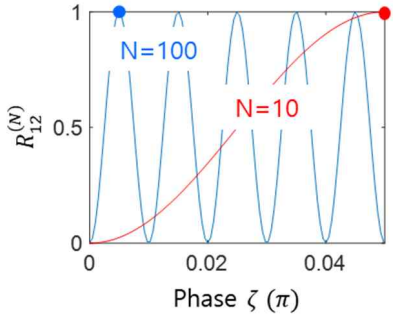

**Figure 3.** Numerical calculations for the normalized Nth-order intensity products using Eq. (9). Blue: N=100, Red: N=10.

Figure 3 shows numerical calculations of Eq. (9), confirming the same superresolution as in Fig. 2, where the resolution enhancement is proportional to the intensity product order N. As discussed in PBWs, thus, the input light at $\lambda_0$ behaves as $\lambda_B = \lambda_0/N$ inside the Sagnac interferometer [14,15,22]. The numerical simulations of Eq. (7) for N=10, 100 are exactly the same as those in Fig. 3 for Eq. (9). Even though the mathematical forms of Eqs. (7) and (9) are different, their quantum behavior of superresolution with near-perfect fringe visibility is equivalent to each other. Thus, Eq. (7) represents the PBW-like superresolution, where the phase quantization with $\xi_k$ is accomplished by the phase controlled polarization-basis projection measurements [29]. As energy of an electron in an atom is quantized by $E_n = nhf$ in the view point of the particle nature of quantum mechanics, the superresolution confirmed in Figs. 2 and 3 via Eqs. (7) and (9), respectively, represents a sort of phase quantization in $\xi_k$, where $k$ is the block number of paired quantum erasers for the projection measurements in Fig. 1.

Unlike coincidence detection in conventional quantum sensing under the particle nature [14,15], thus, the proposed coherent superresolution method is counterintuitively macroscopic and deterministic according to the wave nature of a photon. Here, it should be noted that the particle and wave natures in quantum mechanics are mutually exclusive according to the Copenhagen interpretation [38]. Thus, the intensity-product method between paired phase-controlled SLMs replaces the conventional coincidence detection technique in N00N-based quantum sensing. In RLG based on Sagnac interferometer, ultrahigh sensitivity of $10^{-15}$ rad/s has already been achieved experimentally [39,40]. Consequently, the proposed noise-free Sagnac superresolution technique gives a great advantage to beat even the state-of- art RLG technology in both resolution and phase sensitivity, where the phase sensitivity of PBW-like superresolution is conditional with a near perfect fringe visibility of the superresolution, which has already been experimentally demonstrated in an MZI scheme [26]. The only difference of the proposed Sagnac superresolution from conventional RLGs is to use orthogonally polarized input lights for the polarization-basis projection measurement of the output light. This may envisage a quantum RLG in the future, whose enhanced resolution is the order of N.

**Conclusion**

A scalable, noise-free, macroscopic superresolution method was proposed in a Sagnac interferometer using phase-controlled polarization-basis projection measurements. General solutions of the noise-free Sagnac superresolution were coherently derived from scalable Nth-order intensity products, where the control phases of N lights were set to satisfy equally spaced fringes between their first-order intensity correlations. The use of orthogonally polarized lights was the key to the projection measurement, satisfying the macroscopic quantum eraser. To solve the complexity issue in the linear optics-based superresolution, a pair of SLMs were introduced for the same projection measurements. Finally, analytically derived general solutions were numerically confirmed for the Sagnac superresolution with near-perfect fringe visibility and compared with PBWs in the conventional N00N-based quantum sensing. Thus, the highest order N in the proposed Sagnac superresolution is practically unlimited due to the huge number of photons in the input laser. Compared to the MZI scheme in N00N-based quantum sensing, the proposed Sagnac superresolution is inherently noise free due to the shared optical paths between counterpropagating lights. Compared to the N limited N00N-based quantum sensing far less than 100, the proposed Sagnac superresolution should pave a road to the quantum sensor beating the high-end classical sensors such as RLG based on multi-wave interference. In addition, it was discussed that the ordered intensity products in the Sagnac superresolution could be interpreted as phase quantization in the viewpoint of the wave nature in quantum mechanics, corresponding to the energy quantization in the particle nature.

**Methods**

The polarizing beam splitter (PBS) in Fig. 1 provides a poloarization-basis Sagnac interferometer with orthogonally polarized counterpropagating light fields. The output field of the Sagnac interferferometer is equally divided into K optical blocks (see dotted boxes) for phase controlled projection measurements, satisfying the quantum eraser scheme [29,30]. Random polarization bases of an input photon (light) for PBS is provided by a 22.5-degree rotated half-wave plate. For the projection measurement of the N-divided output fields, 45°-rotated N polarizers are used, resulting in the quantum eraser effect with interference fringes [25,26]. For the out-of-phase phase controls in each block, a birefringent material such as a quarter-wave plate (QWP) is used, where the rotation angle of QWP determines the relative phase shift to the vertically polarized light component [12,25]. For the equal fringe shift among all quantum erasers involved for the intensity product, discrete rotation angles of QWPs are assigned, where equally spaced fringes are formed by the rotation velocity $\Omega$ of the Sagnac interferometer. Each pixel pair of spatial light modulators (SLMs) in the Inset of Fig. 1 is to replace the optical block, solving the complexity issue. The maximum scalability of the intensity product in the Sagnac interferometer is ideally up to the photon number but practically to the pixel number M of the SLM in the Inset of Fig. 1, where the photons of 1 mW HeNe laser is about $10^{15}$/s.

**Data Availability**
All data generated or analyzed during this study are included in this published article.

**Funding:** This work was supported by the IITP-ITRC (IITP-2025-RS2021-II211810) funded by the Korea government (Ministry of Science and ICT).

**Author contribution:** BSH solely wrote the paper.

**Competing Interests:** The author is the founder of Qu-Lidar.